# Chaotic-Based Processor for Communication and Multimedia Applications

Fei Li
09212020027@fudan.edu.cn

Chaos is a phenomenon that attracted much attention in the past ten years. In this paper, we analyze chaos-based signal processing, and proposed a chaos processor to take advantage of chaos phenomenon. We also analyzed and demonstrated two of its practical applications in communication and sound synthesis.

1. **Chaos Processor**

    With the advent of machine learning algorithms [1-7], many people claim artificial intelligence can replace people in daily work, such as face recognition, decision making, and financial activities. However, in many realms, we still need inspiration from nature. Chaos is a phenomenon that attracted much attention in the past ten years [8].

    Chaos phenomenon is usually observed in dynamic systems which are especially sensitive to initial conditions [8]. Different from linear systems, chaos system is an extremely non-linear system where a little change in initial (present) condition can greatly change the output (future).

    Chaos can be used for signal processing. Since little change in initial condition can greatly change output, it is extremely useful for encryption and communication. With chaos signal processing, the initial condition is very hard to be estimated for attackers which only have output condition. This enables efficient encryption and secure communication.

    In order to achieve chaotic signal processing, a dedicated chaos processor is required. The chaos processor is composed of two parts: chaos signal generation and chaos signal retrieval (Fig. 1). Chaos signal generation is used to generate chaos signal, while chaos signal retrieval is used to retrieve initial state of received chaos signal. Since chaos signal is analog in nature, with conventional CPU, it is hard to generate truly chaotic signal, or retrieve initial state of chaotic signal. On the contrary, a chaos processor based on analog signal processing is necessary for efficient chaotic signal processing.

    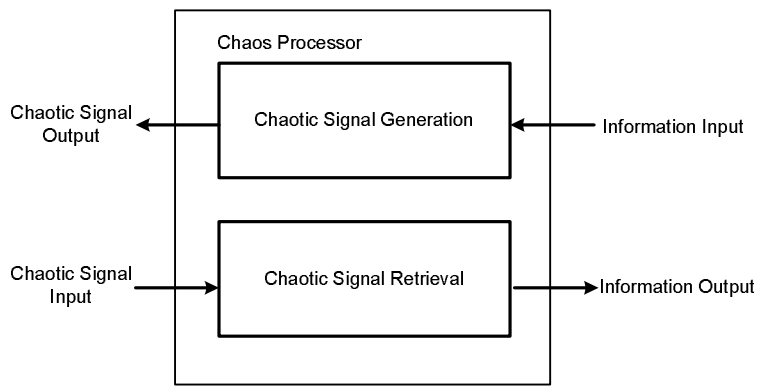

    Fig. 1 Diagram of Chaos Processor

## 2. Chaos Signal Generator and Retriever

Chua's circuit was proposed by L. O. Chua [8], and it quickly drew people's attention after publication. It was the first time that people experimentally observed chaos in real circuit. After that, more than 1000 papers based on Chua's circuit have been published. Until now, Chua's circuit(and its variations) is still an active region for research.

The basic Chua's circuit is plotted in Fig. 2.

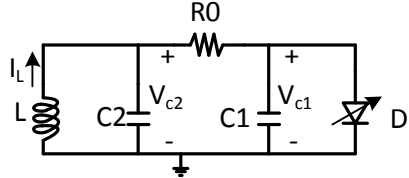

Fig. 2 Basic Chua's circuit

It includes one inductor, one resistor, two capacitors and one non-linear device (Chua's diode), where Chua's diode has piecewise linear current-voltage curve (as plotted in Fig. 2) and is the most important component in this circuit.

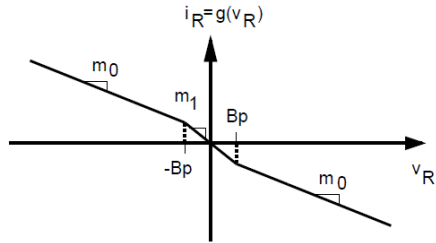

Fig. 2 I-V curve of Chua's diode

Suppose the I-V relationship of Chua's circuit can be expressed with $i_R=g(V_R)$ (as in Fig. 2), then the differential formula for Chua's circuit can be expressed as:

$$C_1 \frac{dV_{c1}}{dt} = \frac{V_{c2} - V_{c1}}{R} - g(V_{c1})$$

$$C_2 \frac{dV_{c2}}{dt} = \frac{V_{c1} - V_{c2}}{R} + I_L \quad (1)$$

$$L \frac{dI_L}{dt} = -V_{c2}$$

The non-linearity of Chua's diode causes the chaotic characteristic of Chua's circuit and its chaotic behavior was mathematically proved in [9].

To implement the characteristic of Chua's diode, operational amplifiers can be used [10]. First, we can analyze the circuit behavior in Fig. 3.

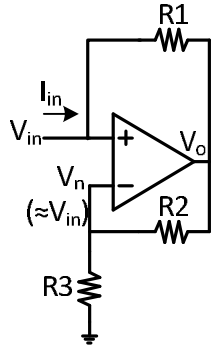

Fig. 3 Basic cell to implement Chua's diode

With high gain of operational amplifier (often greater than 60 dB) and negative feedback, the voltage difference at the plus input ($V_p$) and minus input ($V_n$) is negligible to maintain finite output. Thus, $V_p$ can be regarded to be equal to $V_n$. With the high impedance of the operational amplifier, we can write:

$$I_{in} = \frac{V_{in} - V_o}{R1}$$

$$V_n \approx V_{in} = \frac{R3}{R2+R3} V_o \Rightarrow I_{in} = \frac{-R2}{R1 R3} V_{in} \quad (2)$$

Considering the saturation region limited by plus/minus power supply ($E_{sat}$), the I-V curve of the basic cell is plotted in Fig. 4.

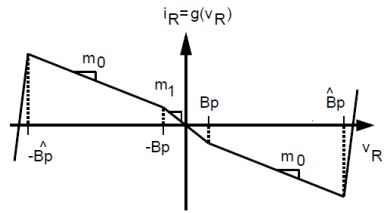

Fig. 4 I-V plot of basic cell for Chua's diode

The slope and knee of curve is controlled by value of resistor R1, R2 and R3. By cascading of those two basic cells (with different slopes and knees by using different resistor values), Chua's diode can be implemented with practical circuit, as shown in Fig. 5.

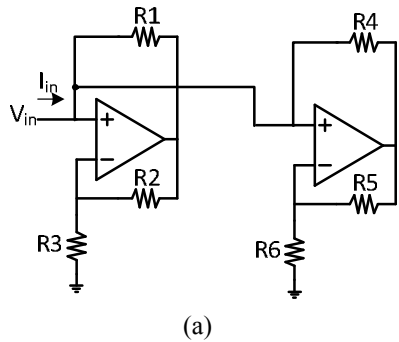

(a)

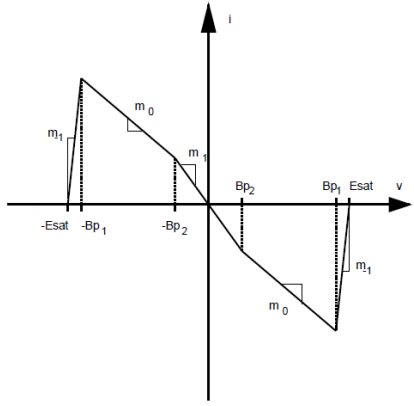

(b)

Fig. 5 (a) Implementation of Chua's diode with operational amplifier and (b) its I-V curve

The implemented Chua's circuit for simulation is shown in Fig. 6.

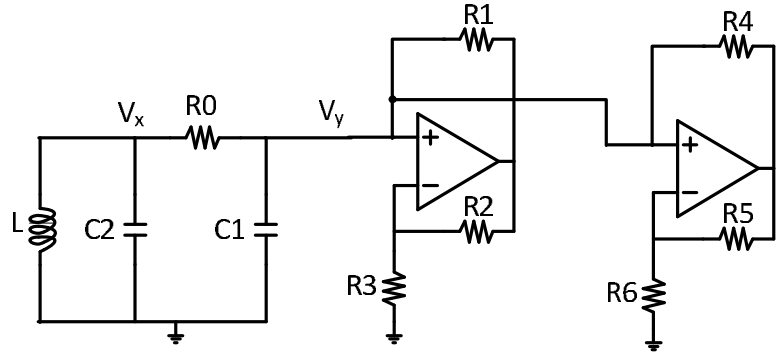

Fig. 6 Implemented Chua's circuit for simulation

3. **Chaotic Characteristic of Chua's Circuit**

In order to analyze the global dynamic behavior of Chua's circuit, we first analyze the circuit behavior with Chua's diode working in different regions [11]. When Chua's diode works in the middle region (with slope of -m1), the differential equations can be expressed as:

$$C_1 \frac{dV_{c1}}{dt} = \frac{V_{c2} - V_{c1}}{R} + m_1 V_{c1}$$

$$C_2 \frac{dV_{c2}}{dt} = \frac{V_{c1} - V_{c2}}{R} + I_L \qquad (3)$$

$$L \frac{dI_L}{dt} = -V_{c2}$$

When Chua's circuit works in the outer region, the equations can be written as (here consider the part at the right half plane; the left half plane has the similar behavior by symmetry):

$$C_1 \frac{dV_{c1}}{dt} = \frac{V_{c2} - V_{c1}}{R} + B_{p2}(m_0 - m_1) - m_0 V_{c1}$$

$$C_2 \frac{dV_{c2}}{dt} = \frac{V_{c1} - V_{c2}}{R} + I_L \qquad (4)$$

$$L\frac{dI_L}{dt} = -V_{c2}$$

When R is very large, the three eigenvalues of Jacobian matrix of (4) are either negative or have negative real part, thus the equilibrium point in outer region is stable. On the other hand, the Jacobian matrix of (3) has unstable eigenvalues, so the equilibrium point of 0 is unstable. This makes the circuit to settle at the outer stable equilibrium point.

When R reduces, the real parts of complex eigenvalues of Jacobian matrix of (4) become negative and they become unstable. The Jacobian matrix of (3) has one unstable real eigenvalue and two stable complex eigenvalues. The dynamic behavior is period-1 limited: the trajectory in phase space will orbit attractor of outer space for one cycle. In time domain, the voltage across the capacitor C1 and C2 will oscillate with one period. When R is reduced further, the system will become period-2 limited, period-4 limited, period-8 limited, and finally the system becomes chaotic: attractor becomes strange attractor (spiral-Chua attractor) the trajectory in phase space orbits the attractor with infinite period. Since the non-linearity of Chua's diode is symmetric, two possible attractors exist in phase space, and the trajectory will orbit either one depends on the initial condition.

When R is reduced even further, two attractors merge and become double scroll. In this case, the trajectory will orbit both attractors. The two attractors in double scroll become closer to each other with lower R. Finally, when the R is so small that the operational amplifiers work in saturation region, the circuit is no longer chaotic but works in saturated region.

Here, we fix the value of L, C1, C2, R1, R2, R3, R4, R5 and R6 in the circuit, and change the value of R0 to observe the behavior of circuit. Spectre® circuit simulation tool is used to simulate the transient waveform. The values of L, C1, C2, R1, R2, R3, R4, R5 and R6 are listed in Table I:

Table I. Values of components in Chua's circuit

|    | Value   |
| --- | --- |
| L  | 18 mH  |
| C1 | 10 nF  |
| C2 | 100 nF |
| R1 | 220 Ω  |
| R2 | 220 Ω  |
| R3 | 2.2 kΩ |
| R4 | 22 kΩ  |
| R5 | 22 kΩ  |
| R6 | 3.3 kΩ |

When R0=2.2 kΩ, the circuit settles in equilibrium point, and no oscillation is observed, as shown in Fig. 7 (waveform of $V_x$):

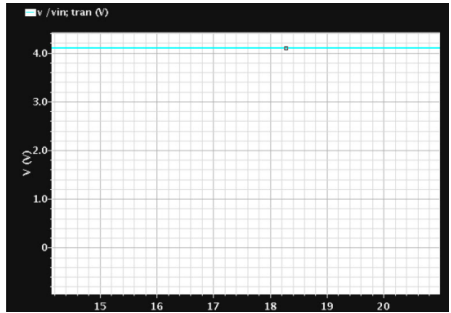

Fig. 7 Transient waveform when R0=2.2 kΩ

When R0=1.9 kΩ, the circuit is period-1 limited, as shown in Fig. 8. The transient waveform has one oscillation cycle, and the trajectory in phase space encircles the attractor once.

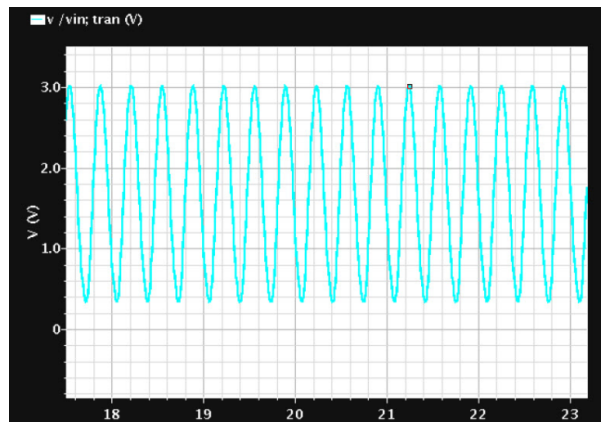

(a)

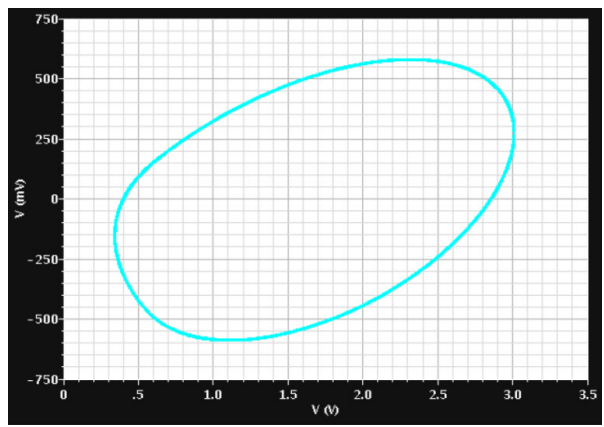

(b)

Fig. 8 Transient waveform (a) and phase space trajectory (b) when R0=1.9 kΩ

When R0=1.87 kΩ, the circuit is period-2 limited, as shown in Fig. 9. The transient waveform has two oscillation cycles, and the trajectory in phase space encircles the attractor twice.

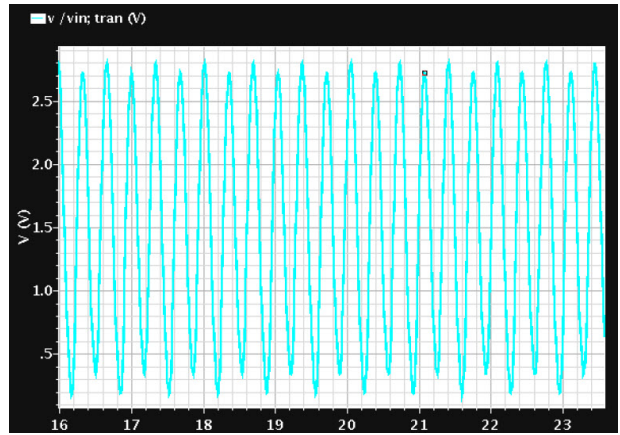

(a)

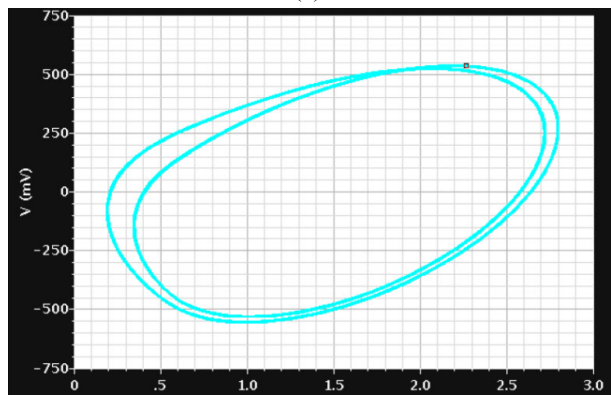

(b)

Fig. 9 Transient waveform (a) and phase space trajectory (b) when R0=1.87 kΩ

When R0=1.85 kΩ, the circuit is period-4 limited, as shown in Fig. 10. The transient waveform has four oscillation cycles, and the trajectory in phase space encircles the attractor for four times.

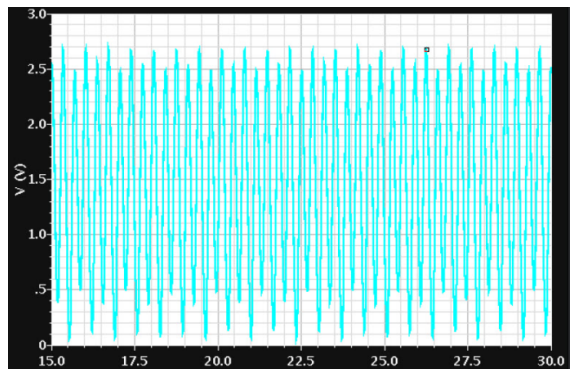

(a)

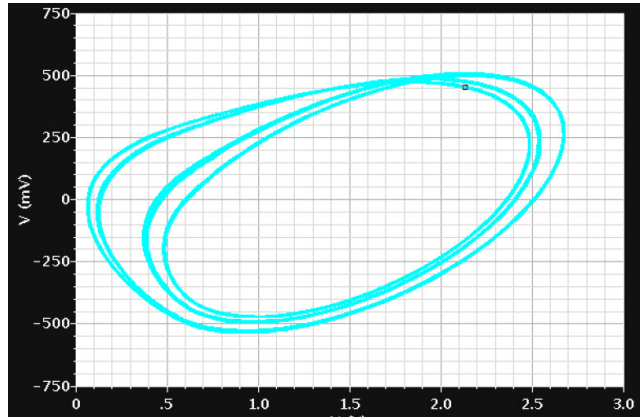

(b)

Fig. 10 Transient waveform (a) and phase space trajectory (b) when R0=1.87 kΩ

When R0=1.8 kΩ, the circuit becomes chaotic (spiral-Chua attractor), as shown in Fig. 11. The transient waveform is no longer periodic, and the trajectory in phase space encircles the attractor for infinite times.

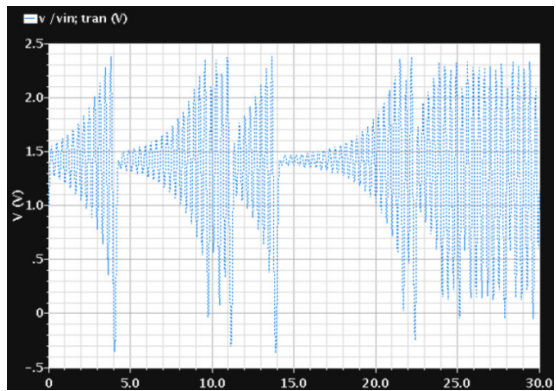

(a)

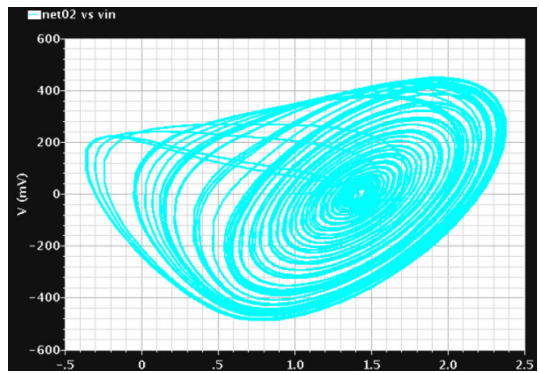

(b)

Fig. 11 Transient waveform (a) and phase space trajectory (b) when R0=1.8 kΩ

When R0=1.7 kΩ, the attractor becomes double-scroll, as shown in Fig. 12. The transient waveforms of $V_x$ (upper part) and $V_y$ (lower part) are shown here.

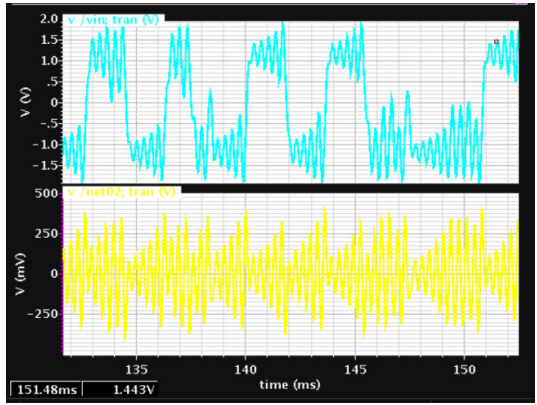

(a)

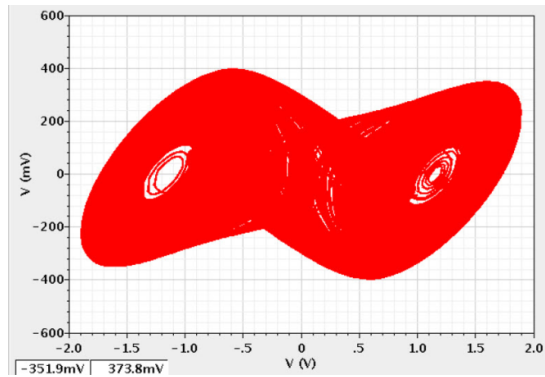

(b)

Fig. 12 Transient waveform (a) and phase space trajectory (b) when R0=1.7 kΩ

When R0 is 1.5 kΩ, the two attractors in double scroll become closer to each other, as shown in Fig. 13.

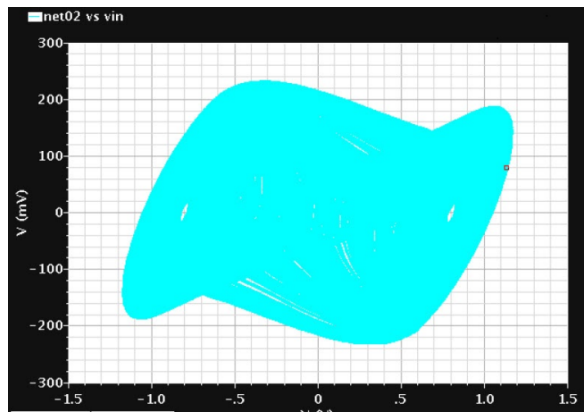

Fig. 13 Trajectory in phase space when R0=1.5 kΩ

When R0 is further reduced to 1 kΩ, the operational amplifier works in saturation region, and the circuit is no longer chaotic. The voltage at $V_x$ and $V_y$ oscillates at one definite period, as shown in Fig. 14. The waveform is not sinusoidal because of the distortion brought by saturation.

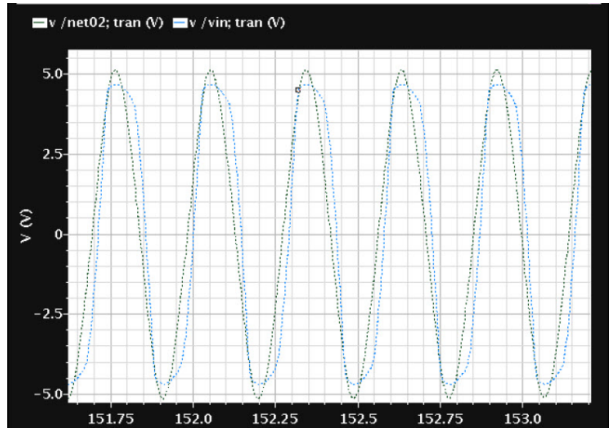

Fig. 14 Waveform when R0=1 kΩ

## 4. Application of Chua's Circuit: Secure Communication

Due to the chaotic nature of Chua's circuit, the output is not predictable without full knowledge of its history. Another interesting characteristic of Chua's circuit is synchronization: two Chua's circuits can be synchronized to have the same output waveform even with different initial conditions. This can be used for secure communication [12].

The scheme for secure communication with Chua's circuit is shown in Fig. 15. The switch is closed when synchronization begins.

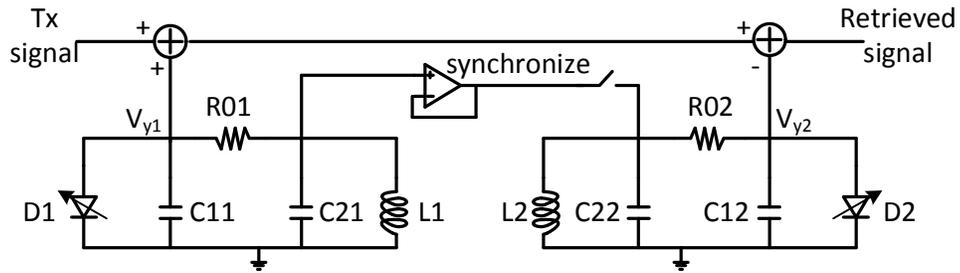

Fig. 15 Secure communication with Chua's circuit

In the first simulation, two Chua's circuits with exactly the same components but different initial conditions are synchronized. The synchronization happens at 100 ms. The waveforms of $V_{y1}$ and $V_{y2}$ (in Fig. 14) before/after synchronization are shown in Fig. 16. The waveform difference is shown in Fig. 17.

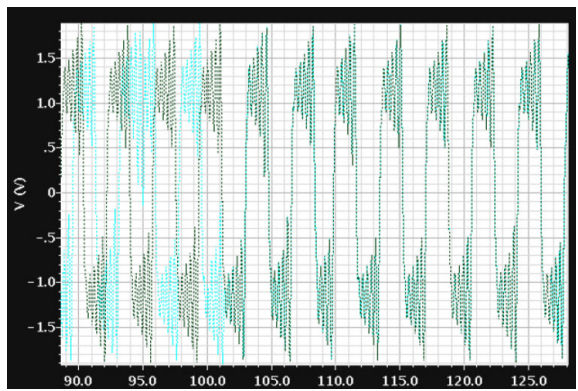

Fig. 16 Waveforms of two Chua's circuits before/after synchonization

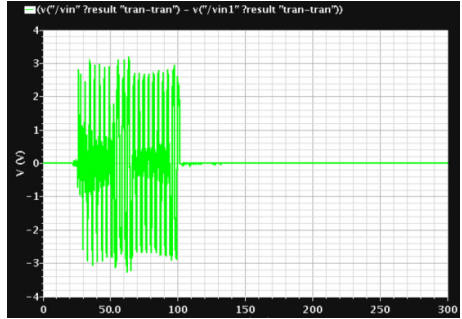

Fig. 17 Difference of $V_{y1}$ and $V_{y2}$ before/after synchronization

It can be seen that before synchronization happens (at 100 ms), two waveforms deviate from each other. After synchronization, two waveforms match with each other and their difference is zero. Thus the signal can be retrieved.

When the two Chua's circuits have some mismatches in components, the synchronization is not perfect. First consider the mismatch in inductor/capacitor. The inductor and capacitor value decides the intrinsic frequency of the oscillation circuit. However, with injection locking phenomenon, two oscillators with different oscillation frequency can be synchronized to the same oscillation condition [13]. Thus, we can expect that it is possible to retrieve signal from a Chua's circuit with small mismatch of L and C. The simulation is run with 5% mismatch of C1 and C2 for two Chua's circuit. The waveforms and difference between $V_{y1}$ and $V_{y2}$ are shown in Fig. 18 and Fig. 19. The synchronization begins at 100 ms.

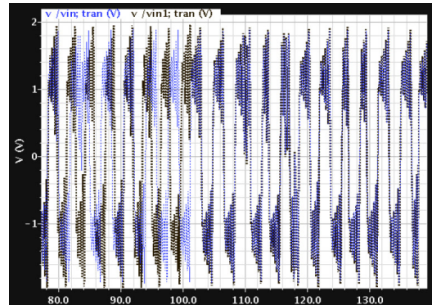

Fig. 18 Waveforms of two Chua's circuit with 5% capacitor mismatch before/after synchronization

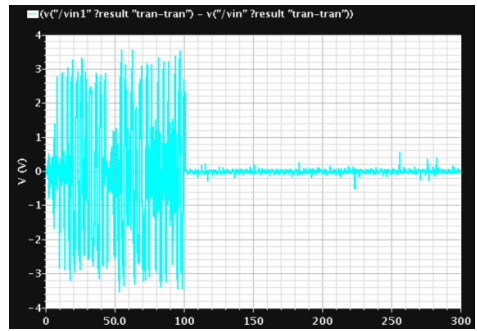

Fig. 19 Difference of two Chua's circuit with 5% capacitor mismatch before/after synchronization

It can be seen that two circuits can be synchronized after 100 ms: though there exists some small "glitches" in the difference of two waveforms, it is possible to retrieve signal since the glitch is small compared with signal amplitude.

On the other hand, the resistor R0 decides the chaotic oscillation attractor of Chua's circuit. Thus, if there is some mismatch in R0, it is possible that those two Chua's circuits cannot be synchronized. Simulation is run with 5% mismatch in R0. The difference of waveforms of $V_{y1}$ and $V_{y2}$ before/after synchronization is shown in Fig. 20. Synchronization happens at 100 ms.

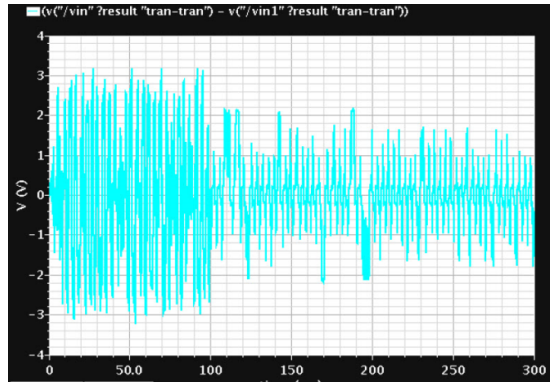

Fig. 20 Difference of $V_{y1}$ and $V_{y2}$

It can be seen from Fig. 20 that the two circuits are not fully synchronized, and it is very hard to retrieve signal with $V_{y2}$.

5. **Application of Chua's Circuit: Sound Synthesis**

Another interesting application of Chua's circuit is sound synthesis [14]. Ordinary oscillator oscillates with only one frequency, and it can only produce one tone. On the other hand, Chua's circuit oscillates with more than one frequency, and its output waveform is more like natural sounds.

The schematic for sound synthesis of Chua's circuit is shown in Fig. 21. The resistor R0 is modulated with certain external signal to change the chaotic behavior of the circuit.

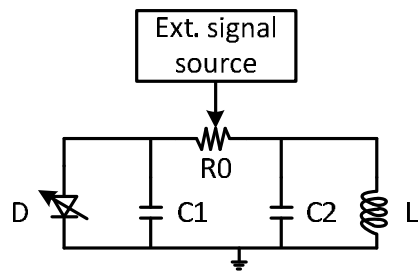

Fig. 21 Schematic of sound synthesis by Chua's circuit

Two simulations were run. In the first simulation, the resistor was modulated with a stair-case wave (with three steps) of 100 Hz. In the second simulation, the resistor was modulated with a sine-wave of 100 Hz. Sample waveform is shown in Fig. 22.

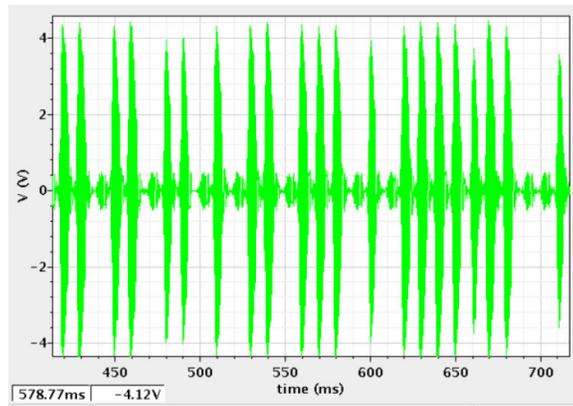

Fig. 22 Sample waveform of sound synthesis by Chua's circuit

The simulated waveform is imported to Matlab and the by using 'sound' function we can hear the corresponding sound of the waveforms. The first waveform sounds like drum, and the second waveform sounds like the sound of insect.

**Conclusion**

In this paper, chaotic behavior and applications of chaos processor are proposed and studied. The behavior of Chua's circuit is strongly associated with the coupling resistor R0 in the circuit, which directly decides the trajectory type in phase space. With different values of R0, the Chua's circuit may work in equilibrium region, period-n limit region, chaotic region or saturation region. In the synchronization of Chua's circuit, the synchronization behavior is more sensitive to R0 mismatch than L and C mismatch, as it decides the chaotic behavior of the circuit. In sound synthesis application, interesting sounds are synthesized with Chua's circuit.